\RequirePackage[2020-02-02]{latexrelease}
\documentclass[aps,prd,groupedaddress,graphicx,nofootinbib]{revtex4}
\usepackage{amsmath,amssymb,graphics,graphicx,color,epsf}
\usepackage{subfigure}
\usepackage[scanall]{psfrag}  
\usepackage{tikz}
\DeclareMathOperator{\sech}{sech}

\begin{document}

\title{On the oscillations of the inflaton field of the simplest $\alpha$-attractor T-model}

\author{Chia-Min Lin}

\affiliation{Fundamental General Education Center, National Chin-Yi University of Technology, Taichung 41170, Taiwan}



\begin{abstract}
In this work, we consider homogeneous oscillations of the inflaton field after inflation. In particular, we obtain an analytical result for the (average) equation of state for the oscillating inflaton field for the simplest $\alpha$-attractor T-model. We also study the post-inflationary evolution before inflaton decay. It is possible that during inflaton field oscillation, the (average) equation of state is that of a cosmological constant. This leads to the phenomenon of oscillating inflation. However, we show that the duration of oscillating inflation is very limited.

\end{abstract}
\maketitle
\large
\baselineskip 18pt
\section{Introduction}

Cosmic inflation \cite{Starobinsky:1980te, Guth:1980zm, Linde:1981mu} is probably the most popular scenario for the very early universe cosmology. 
It can explain many problems of our universe, such as why it is so big and why it is so homogeneous but not that homogeneous. 
In many models of cosmic inflation, inflation is driven by the potential energy density of a scalar field called the inflaton field. 
Apparently, inflation has to end (or have a graceful exit, as is often put). We have to recover a decelerating universe at least in our observable universe in order to have successful structure formation.
After slow-roll inflation, the inflaton field starts to enter an oscillation phase. If we can understand better about this oscillation period, we would have a better understanding of the post-inflationary evolution history such as the early dark age, the mechanism of (p)reheating, and/or baryogenesis to name a few.

The (average) equation of state of the oscillation of a scalar field in an expanding universe \cite{Turner:1983he} is well known for a potential of a power function $V=\lambda \phi^n$. It seems this is the only example where we know a closed-form solution. In this work, we investigate the simplest\footnote{In general, $\alpha$-attractor T-model takes the potential form $V=V_0 \tanh^{2n}\left( \frac{\phi}{F} \right)$, the simplest one corresponds to $n=1$.} $\alpha$-attractor T-model of inflation \cite{Kallosh:2013hoa, Kallosh:2013yoa, Carrasco:2015pla} with the potential
\begin{equation}
V(\phi)=V_0\tanh^2\left( \frac{\phi}{F} \right).
\label{alpha}
\end{equation}
We show that the (average) equation of state can be solved in this model. By using this result, we study the post-inflationary evolution of the simplest $\alpha$-attractor T-model of inflation before the inflaton decays. In the following, I start a journey of this calculation.

\section{oscillating scalar field in an expanding universe}

The oscillation of a homogeneous scalar field in an expanding universe was considered in \cite{Turner:1983he}. We provide a brief review of the relevant parameters in this section and fill in some gaps in the calculations.

For a homogeneous scalar field $\phi$ with potential $V$, the energy density $\rho$ is given by
\begin{equation}
\rho=\frac{\dot{\phi}^2}{2}+V,
\label{rho1}
\end{equation}
and the pressure is
\begin{equation}
p=\frac{\dot{\phi}^2}{2}-V.
\label{p}
\end{equation}
Therefore, for an oscillating scalar field, we have
\begin{equation}
\dot{\phi}^2=\rho+p = \left( 1+\frac{p}{\rho} \right)\rho \equiv \left( 1+w \right) \rho \equiv \left( \gamma+\gamma_p \right) \rho,
\label{gp}
\end{equation}
where $w$ is the parameter for the equation of state. We have defined $\gamma$ as the average of $(\rho+p)$ over an oscillation, and $\gamma_p$ is the periodic part of $\dot{\phi}^2$ following \cite{Turner:1983he}. It is assumed that the frequency of oscillations $\omega \simeq \dot{\phi}/\phi$ satisfies
\begin{equation}
\frac{1}{\omega} \ll \frac{1}{H}.
\label{omega}
\end{equation}
This means the period of oscillation is much smaller than the characteristic time scale (or ``age") of the universe. 

We would like to calculate the (average) equation of state.
In a textbook of cosmology \cite{Mukhanov:2005sc}, the argument goes like this. If we neglect the Hubble expansion (due to the fast oscillation given by Eq.~(\ref{omega})), the equation of motion becomes
\begin{equation}
(\phi\dot{\phi}\dot{)}-\dot{\phi}^2+\phi V^\prime=0.
\end{equation} 
This is nothing but the Klein-Gorden equation for a homogeneous scalar field. The first term drops out upon averaging over one period and gives $\langle \dot{\phi}^2 \rangle = \langle \phi V^\prime \rangle$.
Therefore from Eqs.~(\ref{rho1}) and (\ref{p}), we obtain
\begin{equation}
\langle w \rangle=\frac{p}{\rho}=\frac{\langle \phi V^\prime \rangle -\langle 2V \rangle}{\langle \phi V^\prime \rangle +\langle 2V \rangle}.
\end{equation}
For example, if $V=\lambda \phi^n$, we have
\begin{equation}
\langle w \rangle=\frac{n\langle \phi^n \rangle -2\langle \phi^n \rangle}{n\langle \phi^n \rangle +2\langle \phi^n \rangle}=\frac{n-2}{n+2}.
\end{equation}
Here the $\langle \phi^n \rangle$ cancels out and we do not have to worry about it. This cancellation only happens when $\phi V^\prime \propto V$, namely the power function. Usually, it is argued that for a general potential, one can power expand it and when the field value is small, the lowest power term dominates and we can use the above formula. 
In this work, we would like to be a little bit more ambitious to calculate the (average) equation of state.

Let us return to the treatment of \cite{Turner:1983he}.
From the continuity equation
\begin{equation}
\frac{d\rho}{dt}=-3H(\gamma+\gamma_p)\rho,
\label{con}
\end{equation}
we have
\begin{equation}
\frac{d\rho}{\rho}=-3\frac{da}{a}\gamma  -3H \gamma_p dt.
\label{c}
\end{equation}
When we integrate Eq.~(\ref{c}), the second term on the right-handed side is negligible due to the condition given by Eq.~(\ref{omega}). In particular, if $\gamma$ is constant, we obtain
\begin{equation}
\rho=a^{-3\gamma}.
\end{equation}
This expression is simpler than that by using $\langle w \rangle$. This is one of the reasons to use $\gamma$.
By definition, $\gamma$ is the average value of $1+w$. Let us calculate the average over one period from $t=0$ to $t=T$. During one oscillation, due to Eq.~(\ref{omega}), $\rho$ can be treated as a constant and we can write $\rho=V_m$, with $V_m$ the potential at the maximum value (turning point of the oscillation) of $\phi$ at $\phi_m$ where $\dot{\phi}_m=0$. The average of $1+w$ is
\begin{equation}
\gamma=1+\langle w \rangle= \frac{1}{T}\int^T_0 \frac{\rho+p}{\rho}dt=\frac{1}{T}\int^T_0 \frac{\dot{\phi}^2}{V_m}dt,
\label{g1}
\end{equation}
where we have used Eqs.~(\ref{rho}) and (\ref{p}) to calculate $\rho+p$.
From Eq.~(\ref{rho}), we have
\begin{equation}
\rho=V_m=\frac{\dot{\phi}^2}{2}+V.
\label{m}
\end{equation}
Therefore 
\begin{equation}
\dot{\phi}=\sqrt{2}\sqrt{V_m-V}.
\label{m2}
\end{equation}
The period is given by
\begin{equation}
T=\int^T_0 dt = 2\int^{\phi_m}_0 \frac{d\phi}{\dot{\phi}}= 2 \int^{\phi_m}_0 \frac{d\phi}{\sqrt{2}\sqrt{V_m-V}}, 
\label{t2}
\end{equation}
where we have assumed the potential is an even function $V(\phi)=V(-\phi)$ and set the lower limit of the integral to $\phi=0$. The last equality is from Eq.~(\ref{m2}). Similarly, we can write
\begin{equation}
\int^T_0 \frac{\dot{\phi}^2}{V_m}dt=2 \int^{\phi_m}_0 \frac{\dot{\phi}^2}{V_m}\frac{d\phi}{\dot{\phi}}=\int^{\phi_m}_0 \frac{\sqrt{2}\sqrt{V_m-V}}{V_m}d\phi.
\label{f}
\end{equation}
By using Eqs.~(\ref{g1}), (\ref{t2}), and (\ref{f}), we obtain \cite{Turner:1983he}
\begin{equation}
\gamma=2 \frac{\int^{\phi_m}_0 \left( 1-\frac{V}{V_m} \right)^{1/2}d\phi}{\int^{\phi_m}_0 \left( 1-\frac{V}{V_m} \right)^{-1/2}d\phi}.
\label{gamma}
\end{equation}
For example, if $V=\lambda \phi^n$, we have
\begin{equation}
\gamma=2 \frac{\int^{\phi_m}_0 \left( 1-\frac{\phi^n}{\phi^n_m} \right)^{1/2}d\phi}{\int^{\phi_m}_0 \left( 1-\frac{\phi^n}{\phi^n_m} \right)^{-1/2}d\phi}.
\label{g}
\end{equation}
In \cite{Turner:1983he}, the author only mentioned that it is ``straightforward" to integrate the above integral to give
\begin{equation}
\gamma=\frac{2n}{n+2}.
\label{pow}
\end{equation} 
I would like to be more explicit here for the readers who do not see the straightforwardness. To my knowledge, I have never seen this calculation presented in a textbook or elsewhere. Let us define
\begin{equation}
x \equiv \left( \frac{\phi}{\phi_m} \right)^n.
\end{equation}
The integral in the numerator becomes
\begin{eqnarray}
\int^{\phi_m}_0 \left( 1-\frac{\phi^n}{\phi^n_m} \right)^{1/2}d\phi &=&\frac{\phi_m}{n} \int^1_0 \left(1-x\right)^{1/2}x^{\frac{1-n}{n}}dx  \\ 
                                                                                                     &=&\frac{\phi_m}{n} \frac{\Gamma\left( \frac{1}{n} \right)\Gamma\left( \frac{3}{2} \right)}{\Gamma\left( \frac{1}{n} +\frac{3}{2} \right)},
\end{eqnarray}
where the result is from a beta function. Similarly, the denominator is
\begin{eqnarray}
\int^{\phi_m}_0 \left( 1-\frac{\phi^n}{\phi^n_m} \right)^{-1/2}d\phi &=&\frac{\phi_m}{n} \int^1_0 \left(1-x\right)^{-1/2}x^{\frac{1-n}{n}}dx  \\ 
                                                                                                     &=&\frac{\phi_m}{n} \frac{\Gamma\left( \frac{1}{n} \right)\Gamma\left( \frac{1}{2} \right)}{\Gamma\left( \frac{1}{n} +\frac{1}{2} \right)},
\end{eqnarray}
Therefore we can obtain
\begin{equation}
\gamma=2 \frac{\Gamma\left( \frac{1}{2}+1 \right)\Gamma\left( \frac{1}{n} +\frac{1}{2} \right)}{\Gamma\left( \frac{1}{2} \right)\Gamma\left( \frac{1}{n} +\frac{1}{2}+1 \right)}=\frac{2n}{n+2},
\end{equation}
where I have used $\Gamma(x+1)=x\Gamma(x)$.
This is one method to do the integral, here I have another method to do the integral without using beta functions. Let us write 
\begin{eqnarray}
\int^{\phi_m}_0 \left( 1-\frac{\phi^n}{\phi^n_m} \right)^{1/2}d\phi&=&\int^{\phi_m}_0 \left( 1-\frac{\phi^n}{\phi^n_m} \right)^{-1/2}  \left( 1-\frac{\phi^n}{\phi^n_m} \right)     d\phi  \nonumber  \\
&=&\int^{\phi_m}_0 \left( 1-\frac{\phi^n}{\phi^n_m} \right)^{-1/2} d\phi  -\int^{\phi_m}_0 \frac{\phi^n}{\phi^n_m}\left( 1-\frac{\phi^n}{\phi^n_m} \right)^{-1/2} d\phi \nonumber \\
&=&\int^{\phi_m}_0 \left( 1-\frac{\phi^n}{\phi^n_m} \right)^{-1/2} d\phi  -\int^{\phi_m}_0 n \frac{\phi^{n-1}}{\phi^n_m}\frac{\phi}{n} \left( 1-\frac{\phi^n}{\phi^n_m} \right)^{-1/2} d\phi  \nonumber  \\
&=&\int^{\phi_m}_0 \left( 1-\frac{\phi^n}{\phi^n_m} \right)^{-1/2} d\phi -\frac{2}{n} \int^{\phi_m}_0 \left( 1-\frac{\phi^n}{\phi^n_m} \right)^{1/2}d\phi,
\end{eqnarray}
where we have used integration by parts. Move the last term to the leftmost of the equalities, and the result follows from Eq.~(\ref{g}). I have done the review part, now let us challenge ourselves to calculate something new.

\section{The simplest $\alpha$-attractor T-model}
\begin{figure}[t]
  \centering
\includegraphics[width=0.6\textwidth]{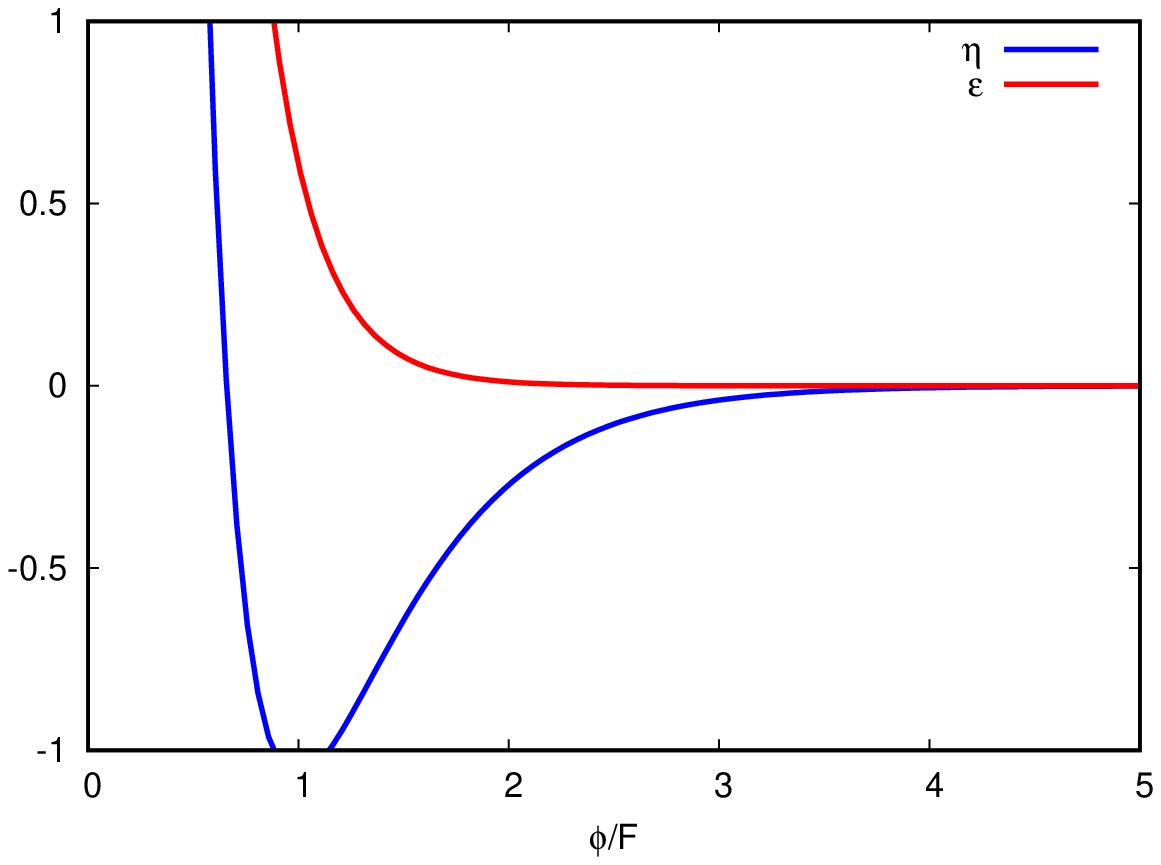}
  \caption{Slow-roll parameters as a function of $\phi/F$ when $F=1$. Slow-roll inflation ends at $\phi_e/F \sim 1$.}
  \label{fig2}
\end{figure}

\begin{figure}[t]
  \centering
\includegraphics[width=0.6\textwidth]{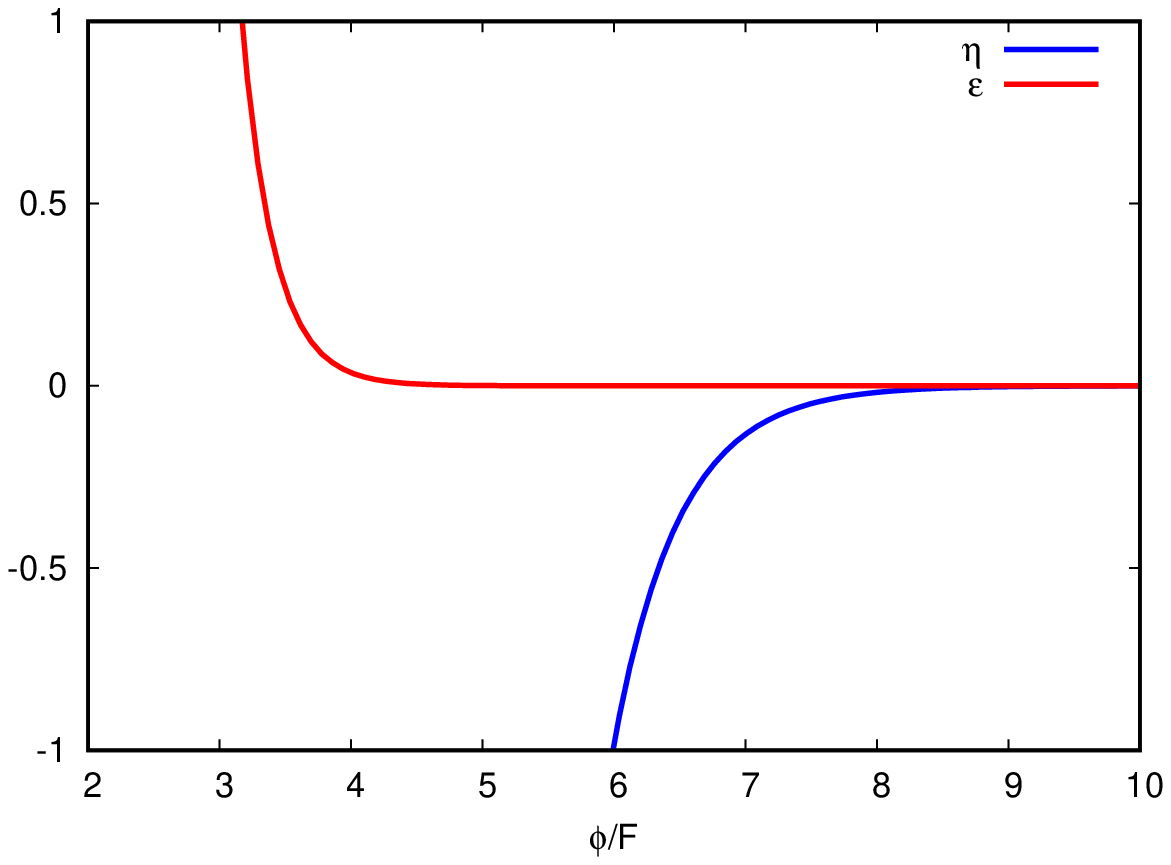}
  \caption{Slow-roll parameters as a function of $\phi/F$ when $F=0.01$. Slow-roll inflation ends at $\phi_e/F \sim 6$.}
  \label{fig3}
\end{figure}

In this section, we do a similar calculation for the potential given by Eq.~(\ref{alpha}).
The first question we need to deal with is perhaps when would (slow-roll) inflation end and the inflaton field starts to oscillate. Let us set the reduced Planck mass $M_P \simeq 2.4 \times 10^{18}\mbox{ GeV}$ to $M_P=1$. We have the slow-rolling parameters (obtained from taking derivatives of Eq.~(\ref{alpha}))
\begin{equation}
\epsilon \equiv \frac{1}{2}\left( \frac{V^\prime}{V} \right)^2=\frac{2}{F^2 \sinh^2\left( \frac{\phi}{F} \right)\cosh^2\left( \frac{\phi}{F} \right)},
\end{equation}
and 
\begin{equation}
\eta \equiv \frac{V^{\prime\prime}}{V}=\frac{2}{F^2\cosh^2\left( \frac{\phi}{F} \right)} \left(  \frac{1}{\sinh^2\left( \frac{\phi}{F} \right)}-2  \right).
\end{equation}
Slow-roll inflation ends at $\phi=\phi_e$ when either $\epsilon=1$ or $|\eta|=1$ is achieved. The slow-roll parameters depend on $F$. We plot the case $F=1$ in Fig.~\ref{fig2} and the case $F=0.01$ in Fig.~\ref{fig3}. As we can see in the figures\footnote{Similar plots for the case which corresponds to $F \sim 1$ can be found in \cite{German:2021tqs}. Our $F$ corresponds to $1/\lambda$ in \cite{German:2021tqs}. Note that we have $M_P=1$.}, slow-roll inflation ends at $\phi_e/F \sim 1$ when $F=1$ and at $\phi_e/F \sim 6$ when $F=0.01$. We assume the inflaton field starts its rapid oscillation and satisfies Eq.~(\ref{omega}) when $\phi_e$ is achieved.

\begin{figure}[t]
  \centering
\includegraphics[width=0.6\textwidth]{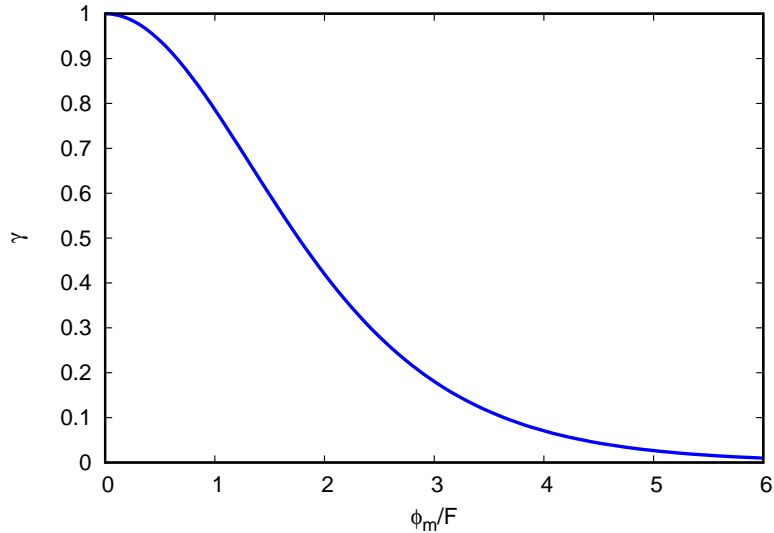}
  \caption{$\gamma$ as a function of $\phi_m/F$ after slow-roll inflation. $\phi_m$ is the envelope of the oscillating $\phi$.}
  \label{fig1}
\end{figure}

Let us calculate $\gamma$ in this model.
For notational simplicity during calculation, we define 
\begin{equation}
c \equiv \frac{V_0}{V_m}=\frac{1}{\tanh^2\left(\frac{\phi_m}{F}\right)}.
\end{equation}
Note that $c>1$. We have identities such as $\sqrt{c-1}=1/\sinh \left(\phi_m/F\right) $.
The numerator of Eq.~(\ref{gamma}) is 
\begin{eqnarray}
\int^{\phi_m}_0 \left( 1-c \tanh^2\left(\frac{\phi}{F}\right) \right)^{1/2}d\phi   
&=&\int^{\phi_m}_0 \left( 1-c \tanh^2\left(\frac{\phi}{F}\right) \right)^{-1/2}d\phi\left( 1-c \tanh^2\left(\frac{\phi}{F}\right) \right)d\phi   \nonumber  \\ &=&\int^{\phi_m}_0 \left( 1-c \tanh^2\left(\frac{\phi}{F}\right) \right)^{-1/2}d\phi \nonumber \\
&-&c \int^{\phi_m}_0 \left(1- \sech^2\left( \frac{\phi}{F} \right) \right)  \left( 1-c \tanh^2\left(\frac{\phi}{F}\right) \right)^{-1/2}d\phi  \nonumber \\
&=&-(c-1)\int^{\phi_m}_0 \left( 1-c \tanh^2\left(\frac{\phi}{F}\right) \right)^{-1/2}d\phi  \nonumber \\
&+&c \int^{\phi_m}_0 \sech^2\left( \frac{\phi}{F}  \right)  \left( 1-c \tanh^2\left(\frac{\phi}{F}\right) \right)^{-1/2}d\phi.  
\label{nu}
\end{eqnarray}
The last integral can be done as
\begin{eqnarray}
c \int^{\phi_m}_0 \sech^2\left( \frac{\phi}{F}  \right)  \left( 1-c \tanh^2\left(\frac{\phi}{F}\right) \right)^{-1/2}d\phi&=&cF \int^{\phi_m}_0 \frac{d\tanh \left( \frac{\phi}{F} \right)}{\sqrt{1-c \tanh^2 \left( \frac{\phi}{F} \right)}} \nonumber \\
&=&cF\left[ \frac{1}{\sqrt{c}} \sin^{-1}\left( \sqrt{c}\tanh\left( \frac{\phi}{F} \right) \right) \right]^{\phi_m}_0 \nonumber \\  &=&\frac{F\pi \sqrt{c}}{2}  \nonumber \\&=&\frac{F\pi}{2 \tanh\left(\frac{\phi_m}{F} \right)}.
\end{eqnarray}
The denominator of Eq.~(\ref{gamma}) is 
\begin{eqnarray}
\int^{\phi_m}_0 \left( 1-c \tanh^2\left(\frac{\phi}{F}\right) \right)^{-1/2}d\phi= F \int^{\phi_m}_0 \cosh \left( \frac{\phi}{F} \right)\frac{1}{\sqrt{1-(c-1)\sinh^2 \left( \frac{\phi}{F} \right)}} d\left( \frac{\phi}{F}\right).
\end{eqnarray}
Let $u=\sqrt{c-1}\sinh \left(\frac{\phi}{F} \right)$, the integral becomes
\begin{eqnarray}
\frac{F}{\sqrt{c-1}} \int^{\sqrt{c-1}\sinh \left(\frac{\phi_m}{F} \right)}_0 \frac{du}{\sqrt{1-u^2}}&=&\left[\frac{F}{\sqrt{c-1}} \sin^{-1}\left( \sqrt{c-1}\sinh \left(\frac{\phi}{F} \right) \right)\right]^{\phi_m}_0 \nonumber \\  &=& \frac{F\pi}{2} \sinh \left( \frac{\phi_m}{F} \right).
\end{eqnarray}
Therefore 
\begin{equation}
\int^{\phi_m}_0 \left( 1-c \tanh^2\left(\frac{\phi}{F}\right) \right)^{1/2}d\phi  =  \frac{F\pi}{2 \tanh\left(\frac{\phi_m}{F} \right)}-\frac{F\pi}{2 \sinh \left(\frac{\phi_m}{F} \right)}.
\end{equation}
This is also the other integral that needs to be done in Eq.~(\ref{nu}).
Substituting these results into Eq.~(\ref{gamma}), we have
\begin{equation}
\gamma=1+\langle w \rangle=\frac{2}{\cosh \left(\frac{\phi_m}{F} \right)+1}.
\label{main}
\end{equation}
This is the main result of our work. We plot $\gamma$ as a function of $\phi_m/F$ in Fig.~\ref{fig1}. Note that for a fixing $F$, $\gamma$ is a function of $\phi_m$, not $\phi$ which is oscillating rapidly. If one studies the equation of state parameter $w$ as a function of $\phi$, it would be oscillating rapidly. Here $\phi_m$ is the envelope of the oscillating $\phi$. When $\phi_m/F \rightarrow 0$, we have $\gamma \rightarrow 1$. This corresponds to $\langle w \rangle =\gamma-1 \rightarrow 0$. The equation of state is that of nonrelativistic (cold) matter. This is reasonable because when $\phi_m/F \rightarrow 0$, the potential in Eq.~(\ref{alpha}) approaches $V \propto \phi^2$. From Eq.~(\ref{pow}), for a quadratic potential $n=2$, we have $\gamma=1$.

On the other hand, when $\phi_m$ becomes large, $\gamma$ decreases toward zero. This can only happen for a small value of $F$. For example, in order to have $\phi_m/F \gtrsim 6$, we need $F \lesssim 0.01$ as can be seen from Fig.~\ref{fig3}. When $\gamma \rightarrow 0$, we have $\langle w \rangle \rightarrow -1$ and the (average) equation of state approaches that of a cosmological constant even for an oscillating scalar field! This phenomenon is known for a different potential and has been called oscillating inflation (or oscillatory inflation) \cite{Damour:1997cb} (see also \cite{Liddle:1998pz, Taruya:1998cz, Cardenas:1999cw, Lee:1999pta, Sami:2001zd, Koutvitsky:2016rkw}). Here we have discovered another example of oscillating inflation\footnote{This phenomenon of having $\langle w \rangle \sim -1$ is also found for the Palatini Higgs inflation with a potential of the form $V=V_0 \tanh^4(\phi/F)$ in \cite{Rubio:2019ypq}.}. The end of slow-roll inflation in this case is the beginning of another period of inflation due to a different mechanism. During oscillating inflation, the inflaton field is not slow-rolling but rapidly oscillating instead. Physically the reason why oscillating inflation can happen is because of the inflaton field spends most of its time on the plateau of the potential during oscillation. In \cite{Kallosh:2013yoa}, the limit of $F \rightarrow 0$ is taken for a universal attractor\footnote{Our $F$ corresponds to $\sqrt{6\alpha}$ of \cite{Kallosh:2013yoa} and $\alpha \rightarrow 0$ is considered.}. Under this condition, oscillating inflation as described above is expected to happen. However, we will show that the period of oscillating inflation in this model is very short in all cases.

\section{Post-inflationary evolution}

In this section, we investigate the post-inflationary evolution of the oscillating inflaton field before inflaton decay.
We would like to know the time dependence of the Hubble parameter, the scale factor, and the amplitude of inflaton field oscillation.
From Eqs.~(\ref{alpha}) and (\ref{m}), we have
\begin{equation}
\rho=V_m=V_0 \tanh^2\left( \frac{\phi_m}{F} \right).
\label{vm}
\end{equation}
The Hubble parameter in a flat universe is given by the Friedmann equation as
\begin{equation}
H=\sqrt{\frac{V_m}{3}}.
\label{h}
\end{equation}
By using Eqs.~(\ref{main}) and (\ref{vm}), we can write
\begin{equation}
\gamma= \frac{2 \left( 1-\frac{V_m}{V_0}\right)^{1/2}}{1+\left( 1-\frac{V_m}{V_0}\right)^{1/2}}.
\label{g}
\end{equation}
By using Eqs.~(\ref{con}), (\ref{vm}), (\ref{h}) and (\ref{g}), we can obtain the average evolution of the energy density as
\begin{equation}
\frac{dV_m}{dt}=-\sqrt{3}V_m^{3/2} \frac{2 \left( 1-\frac{V_m}{V_0}\right)^{1/2}}{1+\left( 1-\frac{V_m}{V_0}\right)^{1/2}}.
\end{equation}
Upon separating variables, we have
\begin{equation}
\frac{1}{2}\left[ V^{-3/2}_m+\frac{1}{V^{3/2}_m\sqrt{1-\frac{V_m}{V_0}}} \right]dV_m=-\sqrt{3}dt.
\end{equation}
Interestingly, this can be integrated to give
\begin{equation}
\frac{1+\sqrt{1-\frac{V_m}{V_0}}}{\sqrt{V_m}}=-\sqrt{3}t,
\end{equation}
where we have absorbed an integration constant into $t$. We can solve $V_m$ as a function of $t$ as
\begin{equation}
\rho=V_m=\frac{12t^2}{\left( 3t^2+\frac{1}{V_0} \right)^2}.
\label{rho}
\end{equation}
From Eq.~(\ref{h}), the Hubble parameter is
\begin{equation}
H=\frac{2t}{3t^2+\frac{1}{V_0}}=\frac{\dot{a}}{a}.
\label{ht}
\end{equation}
From the second equality, the scale factor can be solved to give
\begin{equation}
a = \left( 3t^2+\frac{1}{V_0} \right)^{1/3}.
\label{at}
\end{equation} 
For $t \gg 1/\sqrt{3 V_0}$, we have $H \sim 2/3t$ and $a \sim t^{2/3}$ as expected for a matter-dominated universe. 
From Eqs.~(\ref{vm}) and (\ref{rho}), we have
\begin{equation}
\tanh\left( \frac{\phi_m}{F} \right)=\frac{1}{\sqrt{V_0}}\frac{2\sqrt{3}t}{3t^2+\frac{1}{V_0}},
\end{equation}
which can be rearranged to
\begin{equation}
\frac{\phi_m}{F}= \ln \left( \frac{\sqrt{3V_0}t+1}{\sqrt{3V_0}t-1} \right).
\end{equation}
We can solve the equation for $t$ to obtain
\begin{equation}
t=\frac{1}{\sqrt{3V_0}}\coth \left( \frac{\phi_m}{2F} \right).
\label{t}
\end{equation} 
\section{oscillating inflation?}
In this section, we apply the results of the previous section to study the phenomenon of oscillating inflation.
The end of oscillating inflation occurs at $\gamma=2/3$ (which corresponds to $\langle w \rangle=-1/3$) when $\phi_m=\phi_f$. From Eq.~(\ref{main}), we have
\begin{equation}
\gamma=\frac{2}{3}=\frac{2}{\cosh \left( \frac{\phi_f}{F} \right)+1},
\end{equation}
which can be solved to give
\begin{equation}
\frac{\phi_f}{F}=\ln \left( 2+\sqrt{3} \right) \simeq 1.3.
\label{pf}
\end{equation}

\begin{figure}[t]
  \centering
\includegraphics[width=0.6\textwidth]{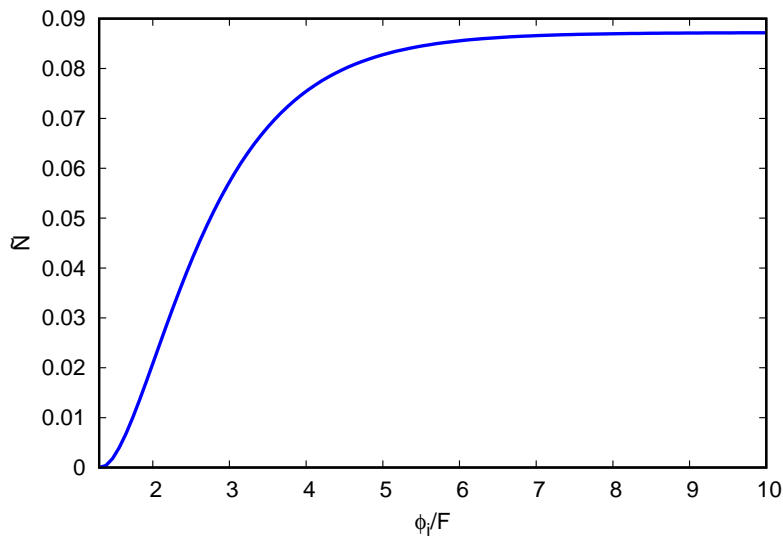}
  \caption{$\tilde{N}$ as a function of $\phi_i/F$. The upper bound is given by $\tilde{N}<0.088$.}
  \label{fig4}
\end{figure}

By using Eqs.~(\ref{ht}), (\ref{at}), and (\ref{t}), we obtain
\begin{equation}
aH=\frac{2t}{\left( 3t^2+\frac{1}{V_0} \right)^{2/3}}=\frac{2V_0^{1/6}}{\sqrt{3}}\frac{\tanh^{1/3} \left( \frac{\phi_m}{2F} \right)}{\left[1+\tanh^2 \left( \frac{\phi_m}{2F} \right) \right]^{2/3}}
\end{equation}
From the beginning of oscillating inflation at $\phi_m=\phi_i$ to the end of oscillating inflation at $\phi_m=\phi_f$,
the relevant number of e-folds is given by \cite{Liddle:1998pz, Liddle:1994dx}
\begin{equation}
\tilde{N}=\ln \frac{a_f H_f}{a_i H_i}=\ln \left(  a_f H_f \times  \frac{\sqrt{3}}{2V_0^{1/6}}  \frac{\left[ 1+\tanh^2 \left( \frac{\phi_i}{2F} \right) \right]^{2/3}}{\tanh^{1/3} \left( \frac{\phi_i}{2F} \right)} \right)<0.088,   
\end{equation}
where $a_f H_f \sqrt{3}/\left( 2V_0^{1/6}\right)=0.688$ can be obtained from Eq.~(\ref{pf}). 
This result is plotted in Fig.~\ref{fig4}.
The upper bound is due to the fact that $\tanh \left( \phi_i/2F \right)<1$. 
Note that this result does not depend on $V_0$ and the upper bound of $\tilde{N}$ holds for any relevant $F$. This shows that $\tilde{N}$ is tiny in any case.
Intuitively, the reason why the value of $\tilde{N}$ is so small may be understood from Eq.~(\ref{vm}). When $\phi_i/F$ is large, the inflaton is near the plateau of the potential. Therefore a small change of $\rho$ leads to a large change in the field value. This can also be seen from Eq.~(\ref{t}).  When $\phi_i/F$ is large, a small change of $t$ corresponds to a large change of the field value. Therefore $\phi_i/F$ cannot be sitting on the plateau for a long time to make a long period of oscillating inflation.

\section{conclusion}
\label{conclu}
In this work, we provide a detailed treatment for an oscillating homogeneous scalar field in an expanding universe. In particular, we find the (average) equation of state for the simplest $\alpha$-attractor T-model given by Eq.~(\ref{main}) which is plotted in Fig.~\ref{fig1}. It shows that when $\phi_m/F \rightarrow 0$, the equation of state is that of nonrelativistic (cold) matter. On the other hand, when $\phi_m/F \gtrsim 6$, which can only happen for $F \lesssim 0.01$, the equation of state approaches that of a cosmological constant. We study the post-inflationary evolution of the model and found an upper bound for $\tilde{N}$ for the duration of the oscillating inflation.

Our result should also have ramifications for the study of (p)reheating \cite{Garcia:2020wiy, Eshaghi:2016kne, Krajewski:2022ezo, Krajewski:2018moi, Ueno:2016dim, German:2020cbw, Shojaee:2020xyr, Podolsky:2005bw, Iarygina:2018kee, Lozanov:2016hid, Li:2020qnk} and oscillons \cite{Copeland:1995fq, Zhang:2020ntm, Lozanov:2017hjm, Lozanov:2016hid, Lozanov:2019ylm, Zhang:2020bec} after $\alpha$-attractor T-model of inflation. 


\acknowledgments
This work is supported by the National Science and Technology Council (NSTC) of Taiwan under grant numbers NSTC 111-2112-M-167-002 and NSTC 112-2112-M-167-001-MY2.

\end{document}